\documentclass[11pt]{article}
\usepackage{amsmath}
\usepackage{amsfonts}
\usepackage{amssymb}
\usepackage{theorem}
\newtheorem{theorem}{Proposition}

\newcommand{\dbar}{\bar{\partial}}
\newcommand{\wt}{\widetilde}
\newcommand{\be}{\begin{equation}}
\newcommand{\ee}{\end{equation}}
\newcommand{\bea}{\begin{eqnarray}}
\newcommand{\eea}{\end{eqnarray}}
\newcommand{\beaa}{\begin{eqnarray*}}
\newcommand{\eeaa}{\end{eqnarray*}}

\begin{document}
\title
{On the $\dbar$-dressing method applicable to heavenly equation}
\author{
L.V. Bogdanov\thanks
{L.D. Landau ITP, Kosygin str. 2,
Moscow 119334, Russia}~
and
B.G. Konopelchenko\thanks
{Dipartimento di Fisica dell' Universit\`a di Lecce
and Sezione INFN, 73100 Lecce, Italy}}
\date{}
\maketitle
\begin{abstract}
The $\dbar$-dressing scheme based on local nonlinear vector $\dbar$-problem
is developed. It is applicable to multidimensional
nonlinear equations for vector fields, and,
after Hamiltonian reduction, to heavenly equation. Hamiltonian reduction is
described explicitely in terms of the $\dbar$-data. An analogue of Hirota
bilinear identity for heavenly equation hierarchy is introduced, $\tau$-function
for the hierarchy is defined. Addition formulae (generating equations) for
the $\tau$-function are found. It is demonstrated that $\tau$-function for
heavenly equation hierarchy is given by the action for $\dbar$-problem
evaluated on the solution of this problem.
\end{abstract}
\section{Introduction}
Dispersionless integrable equations represent themselves an important class
of nonlinear PDEs with various applications in physics. Recently, considerable interest
has been paid to hydrodynamic type equations which arise in the general
relativity, Yang-Mills theory and theory of Einstein-Weyl spaces
(see e.g. \cite{Plebanski}-\cite{dun3}).
Heavenly equations \cite{Plebanski}
have been studied particularly intensively.
Different techniques have been used to analyze
these equations and their properties.

In the present paper we apply the $\dbar$-dressing method to such equations.
The $\dbar$-dressing method has been proposed in \cite{dbar1} and it is applicable
to various classes of PDEs (see e.g. \cite{dbar2,dbar3}). During the last five years
it was demonstrated that the quasiclassical $\dbar$-dressing method is an effective
tool to study dispersionless integrable equations (see e.g. \cite{KMR}-\cite{dconstr}).
Here we will show that the $\dbar$-dressing method based on local nonlinear vector
$\dbar$-problem effectively works for class of equations for vector fields
including heavenly equations. Hamiltonian $\dbar$-problem generating the
heavenly equations is of particular interest.

\section{Basic dressing scheme. Integrable equations for vector fields.}
We start with nonlinear local vector $\dbar$-problem
\be
\dbar S^i=W^i(z,\bar z;S^1,\dots,S^N),\qquad i=1,\dots, N,
\label{dbar}
\ee
where $z\in\mathbb{C}$, bar means complex conjugation,
$
\dbar=\frac{\partial}{\partial \bar z},
$
and functions $W^i$ ($\dbar$-data) are smooth functions
of $S^i$. Similar to standard $\dbar$-dressing method \cite{dbar1},
we suggest that $\dbar$-data
$W^i(z,\bar z;S^1,\dots,S^N)$ are equal to zero in some domain
(or set of domains) $G$ of the complex plane,
thus $\dbar$-data are localized on a closed subset $\mathbb{C}\setminus G$.
This subset may consist
of a number of curves, thus giving rise to vector nonlocal Riemann problem. In the following
we will usually suggest that $\mathbb{C}\setminus G$ is a unit disc $D$.

In the simplest case $N=1$ equation (\ref{dbar}) defines the so-called generalized
analytic functions (see e.g. \cite{Vekua}). Various classes of elliptic systems of the type
(\ref{dbar}) are also studied quite well (see e.g. \cite{Wendland},\cite{Giebert}).

The $\dbar$-dressing method starts with specification of analytic properties of solutions
of the system (\ref{dbar}). Namely,
the problem is to find the vector function $\mathbf{S}=(S^1,\dots,S^N)$ of the form
$\mathbf{S}=\mathbf{S}_0+\mathbf{\wt S}$, satisfying
(\ref{dbar}) in the unit disc, with $\mathbf{\wt S}$ analytic outside the unit disc
and decreasing at infinity, and $\mathbf{S}_0$ analytic in the unit disc,
for arbitrary $\mathbf{S}_0$
($\mathbf{S}_0$ plays the role of the boundary condition). The function $\mathbf{S}$ in this
case may be considered as a functional of $\mathbf{S}_0$.
Introducing a parameterization of $\mathbf{S}_0$,
\be
S^i_0=\sum_{n=1}^\infty t^i_n z^{n-1},
\label{param}
\ee
we may also consider $\mathbf{S}$ as a function of $N$ infinite
sets of times $(\mathbf{t}^1,\dots, \mathbf{t}^N)$.

First derivatives of $\mathbf{S}$ over
times satisfy a linear system
\be
\dbar f^i=\sum_jW^i_{,j}(z,\bar z;S^1,\dots,S^N) f^j,
\label{dbarlin}
\ee
where $W^i_{,j}=\frac{\partial W^i}{\partial S^j}$.
The linear space of functions,
defined by this equation, admits multiplication by arbitrary scalar function
of times and by arbitrary scalar functions of $z$, analytic inside the unit disc.

Solutions of equation (\ref{dbarlin}) which are bounded on the whole complex plane
and decrease at infinity play a special role. In the case $N=1$ there is an analogue
of the Liouville theorem (see e.g. Theorem 3.11 from \cite{Vekua}) which states
that a solution of (\ref{dbarlin}) continuous and bounded on the whole complex plane
and vanishing at some fixed point (in particular, $z=\infty$) is identically zero.
For systems (\ref{dbarlin}) with $N>1$ this is in general not true
(see e.g. \cite{Wendland},\cite{Giebert}). In our approach we assume the validity
of an analogue of Liouville theorem. It is obviously valid for the systems (\ref{dbarlin})
with small $W^i_{,j}$.

However, below we will demonstrate that for some special
reductions of the problem (\ref{dbar}) it is possible to prove an analogue of
Liouville theorem for systems (\ref{dbarlin}) for {\em arbitrary} $\dbar$-data
satisfying the reduction (given the existence of the basis of solutions).

Using the properties of the linear space of solutions of the system (\ref{dbarlin})
and the standard ideology
of the $\dbar$-dressing method \cite{dbar1,dbar2,dbar3},
it is possible to demonstrate that the function $\mathbf{S}$ satisfies an (infinite) set of
linear equations. Compatibility condition for a pair of such equations defines a closed
system of  multidimensional nonlinear PDEs for the coefficients of linear operators.
The first linear equations are
\be
\partial^i_2\mathbf{S}-z\partial^i_1\mathbf{S}=
\sum_{p=1}^N (\partial^i_1u^p)\partial^p_1\mathbf{S},
\label{linear}
\ee
where $\partial^i_n=\frac{\partial}{\partial t^i_n}$, functions
$u^p(\mathbf{t}^1,\dots, \mathbf{t}^N)$ represent first coefficients of expansion of
functions $\wt S^p(z;(\mathbf{t}^1,\dots, \mathbf{t}^N))$ at infinity;
$\wt S^p=\sum_{n=1}^{\infty} \wt S^p_n z^{-n}$, $u^p=-\wt S^p_1$.
Taking compatibility condition
of a pair of linear equations (\ref{linear}) labelled by superscripts $i$ and $j$, we get a closed
equation for the vector field $\mathbf{u}=\sum_{p=1}^N u^p\partial^p_1$,
\be
\partial^i_2\partial^j_1 \mathbf{u}-\partial^j_2\partial^i_1 \mathbf{u}
-[\partial^j_1 \mathbf{u},\partial^i_1 \mathbf{u}]=0,
\label{vector}
\ee
where $[\mathbf{u},\mathbf{v}]$ is a standard commutator of vector fields.
In a similar manner one can construct higher linear problems (\ref{linear})
and nonlinear systems (\ref{vector}).

Lax pairs of the type (\ref{linear}) and corresponding systems (\ref{vector})
are known for a long time (see e.g. \cite{ZS2}). The method to solve
Cauchy problem and the dressing method for (\ref{vector}) based on the
spectral theory of operators of the type (\ref{linear}) has been recently
developed by S.V. Manakov and P.M. Santini \cite{MS}.

In the simplest case $N=1$ our scheme provides us with the $\dbar$-dressing,
linear problems and nonlinear equations for the universal hierarchy of hydrodynamic type
studied in \cite{Alonso,Alonso1}. $\dbar$-dressing for this hierarchy will be studied elsewhere.
In this paper we will concentrate on the Hamiltonian reductions of the equations for vector
fields and associated heavenly type equations.
\section{Heavenly equation}
Let us consider two-component case ($N=2$) and denote $x=t^1_1$, $y=t^2_1$,
$t=t^1_2$, $\tilde t=t^2_2$. It is easy to check that the system (\ref{vector})
admits the reduction to
Hamiltonian
vector field $\mathbf{u}=\Theta_y\partial_x-\Theta_x\partial_y$, for which
it becomes
a second heavenly equation \cite{Plebanski}
\be
\Theta_{ty}-\Theta_{\tilde t x}-\Theta_{xy}^2+\Theta_{xx}\Theta_{yy}=0.
\label{heavenly}
\ee
Hamiltonian reduction of the system (\ref{vector}) surprisingly corresponds to the
Hamiltonian reduction of the $\dbar$-system (\ref{dbar}).
Characterization of this reduction in terms of problem (\ref{dbar}) is given by the
following statements (for simplicity we formulate them for the two-component case,
$N$-component case is completely analogous).
\begin{theorem} Let the $\dbar$-data for the problem (\ref{dbar}) be of the form
\be
W^1=W_{,2},\quad W^2=-W_{,1},
\label{hamred}
\ee
where $W(z,\bar z;S^1,S^2)$ is some function (potential for the $\dbar$-data). Then the
two-forms
\be
w=\delta S^1\wedge\delta S^2,
\label{two-form}
\ee
or,equivalently,
\be
\wt w(\delta\mathbf{S},\tilde\delta{\mathbf{S}}):=
\delta S^1\tilde\delta S^2-\tilde\delta S^1\delta S^2,
\label{two-form1}
\ee
where $\delta$ and $\tilde\delta$ denote arbitrary variations,
are analytic
inside the unit disc (in general case in $\mathbb{C}\setminus G$, which is the
support of
$\dbar$-data).
\end{theorem}
\textbf{Proof.} The problem (\ref{dbar}) in this case looks like Hamilton equations with
complex time $\bar z$ and Hamiltonian $W$,
\be
\dbar S^1=\frac{\partial W}{\partial S^2},\qquad
\dbar S^2=-\frac{\partial W}{\partial S^1},
\label{2dbar}
\ee
and in complete analogy with standard Hamiltonian mechanics, we get
(analogue of Liouville theorem, see e.g. \cite{Arnold})
\be
\dbar(\delta S^1\wedge\delta S^2)=0.
\label{HirotaHE}
\ee
It is also a simple check that
\be
\dbar(\delta S^1\tilde\delta S^2-\tilde\delta S^1\delta S^2)=0.
\label{HirotaHE1}
\ee
\hfill$\square$\\

Analyticity of the two-form $\tilde w$ (\ref{two-form1}) inside the unit disc readily implies
that vector field $\wt S^1_1\partial_x+ \wt S^2_1 \partial_y$ is Hamiltonian.
\begin{theorem}
Let the $\dbar$-problem (\ref{dbar}) be of the Hamiltonian form (\ref{2dbar}).
Then the
vector field $\wt S^1_1\partial_x+ \wt S^2_1 \partial_y$ is Hamiltonian.
\end {theorem}
\textbf{Proof}
Identity (\ref{HirotaHE1}) means that for any pair of solutions
$\mathbf{f}$, $\tilde{\mathbf{f}}$
of linear equations (\ref{dbarlin}) inside the unit disc one has
\be
\dbar(f^1\tilde f^2-\tilde f^1 f^2)=0.
\label{HirotaHE2}
\ee
Taking a pair of solutions $\mathbf{f}=(S^1_x,S^2_x)$,
$\mathbf{\wt f}=(S^1_y,S^2_y)$ and considering
analytic properties of the form (\ref{two-form1}),
we come to the conclusion that
\be
\{S_1,S_2\}:=S^1_xS^2_y-S^1_yS^2_x=1.
\label{bracket}
\ee
The fist term of expansion of this relation at $z\rightarrow\infty$ gives the identity
$\partial_x \wt S^1_1+\partial_y \wt S^2_1=0$. Thus $\wt S^1_1=-\Theta_y$,
$\wt S^2_1=\Theta_x$,
meaning that the vector field
$\wt S^1_1\partial_x+ \wt S^2_1 \partial_y$ is indeed Hamiltonian.\hfill$\square$\\

Identity (\ref{HirotaHE2}) can be used in many different ways. We will show now that
it directly leads to the linear problems of the type (\ref{linear}). Indeed, let us chose
$\mathbf{f}=\mathbf{S}_x$ and $\tilde{\mathbf{f}}=\mathcal{L} \mathbf{S}$, where
$\mathcal{L}$ is a first order differential operator in independent variables depending
on $z$ such that $(\mathcal{L}\mathbf{S})(z)$ is bounded outside the unit disc  and
$(\mathcal{L}\mathbf{S})(z)\rightarrow 0$ as $z\rightarrow 0$. Then due to (\ref{HirotaHE2})
$\tilde w(\mathbf{S}_x, \mathcal{L} \mathbf{S})$ is analytic in the whole complex plane
and vanishes at $z\rightarrow \infty$. Hence $\tilde w(\mathbf{S}_x, \mathcal{L} \mathbf{S})=0$
and consequently $\mathcal{L} \mathbf{S}-\alpha\mathbf{S}_x=0$,
where $\alpha$ is some function of independent variables and $z$.
On the other hand,
choosing $\mathbf{f}=\mathbf{S}_y$ and $\tilde{\mathbf{f}}=\mathcal{L} \mathbf{S}$,
we come to the conclusion that $\mathcal{L} \mathbf{S}-\beta\mathbf{S}_x=0$. Then it
is easy to see that $\alpha=\beta=0$ and $\mathcal{L} \mathbf{S}=0$. So we have obtained
linear problem for $\mathbf{S}$.

For the heavenly equation (\ref{heavenly}) one gets the known linear problems
(see e.g. \cite{Husain})
\be
\begin{array}{l}
\mathbf{S}_{t}-z\mathbf{S}_{x}-\{\mathbf{S},\Theta_{x}\}=0,\\
\mathbf{S}_{\tilde t}-z\mathbf{S}_{y}-\{\mathbf{S},\Theta_{y}\}=0.
\end{array}
\label{linearHE}
\ee

Let us emphasize that our derivation of linear problems in the Hamiltonian case
(\ref{2dbar}) was done for {\em arbitrary} $\dbar$-data (the only thing we need
is {\em existence} of solution), and we didn't suggest validity of an analogue
of Liouville theorem for the system (\ref{dbarlin}) corresponding to Hamiltonian
case (in fact in the process of derivation we have proved it).

\section{$\tau$-function and addition formulae for the heavenly equation hierarchy}
Analyticity of the two-form $w$ (\ref{two-form}) inside the unit disc
or equation (\ref{HirotaHE2})
play a role of Hirota identity for heavenly equation hierarchy. This property
can be formulated in a standard way for the boundary value of $w$ on the unit circle,
using a projection operator. It implies the existence of the $\tau$-function for
the heavenly equation hierarchy (which coincides with $\Theta$ introduced above for the heavenly
equation) and gives `addition formulae' (generating equations) of the hierarchy
and heavenly equation itself.
\begin{theorem}
Identity (\ref{HirotaHE}) implies that one-form
\be
\theta=\frac{1}{2\pi\mathrm{i}}\oint(\wt S^2\delta S^1_0-\wt S^1 \delta S^2_0)dz
\label{1-form}
\ee
is closed.
\end{theorem}
The proof is by simple direct calculation.

We define a $\tau$-function $\Theta(\mathbf{t}_1,\mathbf{t}_2)$
for heavenly equation hierarchy
through closed one-form (\ref{1-form}) by the relation $\delta \Theta=\theta$. Introducing
vertex operators $D^1(z)=\sum_{n=1}^\infty z^{-n}\partial^1_n$,
$D^2(z)=\sum_{n=1}^\infty z^{-n}\partial^2_n$, it is easy to demonstrate that
\be
\wt S^1(z)=-D^2(z)\Theta, \quad \wt S^2(z)=D^1(z)\Theta.
\label{repr}
\ee
Substituting this representation into (\ref{bracket}), we get the equation
\be
D^2(z)\Theta_x-D^1(z)\Theta_y-\{D^1(z)\Theta,D^2(z)\Theta\}=0
\label{heav1}
\ee
The first nontrivial order of expansion of this equation at $z\rightarrow \infty$
gives exactly the heavenly equation (\ref{heavenly}).

To derive addition formulae (generating equations in terms of vertex operators) for $\Theta$,
we will consider the two-forms
{\small $\wt w(D^1(z')\mathbf{S}(z),D^1(z'')\mathbf{S}(z))$,
$\wt w(D^2(z')\mathbf{S}(z),D^2(z'')\mathbf{S}(z))$,
$\wt w(D^1(z')\mathbf{S}(z),D^2(z'')\mathbf{S}(z))$}.
Taking into account the identity (\ref{HirotaHE2}) and analytic properties of these forms,
one gets
{
\begin{multline*}
D^1(z')S^1(z)\cdot D^1(z'')S^2(z)-D^1(z'')S^1(z)\cdot D^1(z')S^2(z)
\\=\frac{1}{z'-z}D^1(z'')\tilde S^2(z')-\frac{1}{z''-z}D^1(z')\tilde S^2(z''),\qquad
\end{multline*}}
\begin{multline}
D^2(z')S^1(z)\cdot D^2(z'')S^2(z)-D^2(z'')S^1(z)\cdot D^2(z')S^2(z)
\\=\frac{1}{z''-z}D^2(z')\tilde S^1(z'')-
\frac{1}{z'-z}D^2(z'')\tilde S^1(z'),
\label{add0}
\end{multline}
\begin{multline*}
D^1(z')S^1(z)\cdot D^2(z'')S^2(z)-D^2(z'')S^1(z)\cdot D^1(z')S^2(z)
\\=
\frac{1}{z'-z}\frac{1}{z''-z}+\frac{1}{z'-z}D^2(z'')\tilde S^2(z')+
\frac{1}{z''-z}D^1(z')\tilde S^2(z'').\qquad
\end{multline*}
These relations directly imply the {\em existence} of the $\tau$-function
(that is equivalent to the definition
in terms of closed variational one-form (\ref{1-form})) and provide us with
addition formulae for it.
Indeed, taking into account
the asymptotic behaviour of both sides of relations (\ref{add0})
(or using the analytic properties of the forms $\wt w$ in the l.h.s.
of these relations and Cauchy formula for the domain $G$),
one gets
\begin{gather*}
D^1(z'')S^2(z')-D^1(z')S^2(z'')=0,\\
D^2(z'')S^1(z')-D^2(z')S^1(z'')=0,\\
D^2(z'')S^2(z')+D^1(z')S^1(z'')=0,
\end{gather*}
that implies (\ref{repr}).

Substituting representation (\ref{repr}) to relations (\ref{add0}),
we obtain a set
of addition formulae
\begin{multline*}
\frac{1}{z'-z}D^1(z'')(D^1(z')-D^1(z))\Theta-
\frac{1}{z''-z}D^1(z')(D^1(z'')-D^1(z))\Theta
\\
=D^1(z'')D^2(z)\Theta\cdot D^1(z')D^1(z)\Theta
-D^1(z'')D^1(z)\Theta\cdot D^1(z')D^2(z)\Theta,
\qquad
\end{multline*}
\begin{multline}
\frac{1}{z''-z}D^2(z')(D^2(z'')-D^2(z))\Theta-
\frac{1}{z'-z}D^2(z'')(D^2(z')-D^2(z))\Theta
\\
=D^2(z')D^2(z)\Theta\cdot D^2(z'')D^1(z)\Theta
-D^2(z')D^1(z)\Theta\cdot D^2(z'')D^2(z)\Theta,
\label{add}
\end{multline}
\begin{multline*}
\frac{1}{z'-z}D^2(z'')(D^1(z')-D^1(z))\Theta-\frac{1}{z''-z}D^1(z')(D^2(z'')-D^2(z))\Theta\\
=D^1(z')D^1(z)\Theta\cdot D^2(z'')D^2(z)\Theta-D^1(z')D^2(z)\Theta\cdot D^2(z'')D^1(z)\Theta.
\qquad
\end{multline*}
Expansion of these equations into powers of parameters $z$, $z''$, $z''$ generates
heavenly equation hierarchy.
\section{$\dbar$-dressing method and $\tau$-function for the heavenly equation hierarchy}
Similar to the case of dispersionless integrable hierarchies \cite{dtau,dconstr},
it is possible
to obtain \textit{explicit} formula for the $\tau$-function of heavenly equation
hierarchy, which is given by the action for the system (\ref{2dbar}) evaluated
on the solution of this system.

The problem (\ref{2dbar}) can be obtained by variation of the action
\bea
f
=\frac{1}{2\pi\text{i}}\iint_{\mathbb{C}\setminus G}
\left(\wt S^2 \dbar \wt S^1 -
W(z,\bar z,S^1,S^2)\right)dz\wedge d\bar z,
\label{action}
\eea
where one should consider independent variations of $\wt{\mathbf{S}}$,
possessing required analytic properties (analytic in
$G$, decreasing at infinity),
keeping $\mathbf{S}_0$ fixed.
\begin{theorem} The function
\be
\Theta(\mathbf{t})=
\frac{1}{2\pi\text{i}}\iint_{D}
\left(\wt S^2(\mathbf{t}) \dbar \wt S^1(\mathbf{t}) -
W(z,\bar z,S^1(\mathbf{t}),S^2(\mathbf{t}))\right)dz\wedge d\bar z,
\label{HEtau}
\ee
where $\mathbf{t}=(\mathbf{t}_1,\mathbf{t}_2)$, $D$ is a unit disc,
i.e., the action (\ref{action}) evaluated on the solution
of the problem (\ref{2dbar}),
is a $\tau$-function of the heavenly equation
hierarchy.
\end{theorem}
\textbf{Proof.} Considering $\Theta(\mathbf{t})$ as a functional of $\mathbf{S}_0$
and calculating its variation, we get
$$
\delta \Theta=\frac{1}{2\pi\text{i}}\oint(\wt S^2\delta S^1_0-\wt S^1 \delta S^2_0)dz,
$$
that coincides exactly with one-form (\ref{1-form}) used to define the $\tau$-function.
\hfill$\square$

Let us consider also the action
\bea
\tilde f
=\frac{1}{2\pi\text{i}}\iint_{D}
\left(S^2 \dbar S^1 -
W(z,\bar z,S^1,S^2)\right)dz\wedge d\bar z,
\label{action1}
\eea
which is an exact analogue of the classical action associated with the Hamiltonian equations
(\ref{2dbar}).
Then
$$
\tilde f=\frac{1}{2\pi\text{i}}\oint S_0^2\wt S^1 dz + f.
$$
So on the solutions of the $\dbar$-problem (\ref{2dbar})
one has ($\wt \Theta=\tilde f$)
$$
\wt \Theta=\Theta-\sum_{n=1}^\infty t^2_n\frac{\partial\Theta}{\partial t^2_n}.
$$
Similar to dispersionless case \cite{dtau,dconstr}, the formula (\ref{HEtau})
provides us with infinitesimal symmetries for the addition formulae (\ref{add})
and in particular for the heavenly equation (\ref{heavenly}). They are given
by the expression
\be
\delta\Theta=-\frac{1}{2\pi\text{i}}\iint_{D}
\delta W(z,\bar z,\mathbf{S})dz\wedge d\bar z,
\ee
where $\delta W(z,\bar z,\mathbf{S})$ is an arbitrary function in the unit disc.

In the simplest case $\delta W=\delta(z-z_0)F(\mathbf{S}(z_0,\mathbf{t}))$,
where $F$ is an arbitrary function, one has
\be
\delta\Theta=F(\mathbf{S}(z_0,\mathbf{t})).
\ee
It is a simple direct check that an arbitrary function $F(S^1,S^2)$ of the solutions
$S^1$ and $S^2$ of linear problems (\ref{linearHE}) solves the linear equation
$$
(\delta\Theta)_{ty}-(\delta\Theta)_{\tilde t x}+\{\Theta_x,(\delta\Theta)_y\}-
\{\Theta_y,(\delta\Theta)_x\}
=0,
$$
which defines infinitesimal symmetries of the heavenly equation (\ref{heavenly}).

Finally we note that, similar to the classical mechanics, canonical transformations
$S^i\rightarrow {S'}^i=f^i(S^1,S^2)$ preserve the two-form  $w$ (\ref{two-form})
and leave $\dbar$-problem (\ref{2dbar}) invariant. In addition,
associated linear problems $\mathcal{L}\mathbf{S}=0$ are invariant
under these transformations.

Under canonical transformation with the generating function $F$ one has
$W\rightarrow W'=W+F_{\bar z}$. So for canonical transformation independent
of $\bar z$ the $\wt\Theta$-function corresponding to (\ref{action1})
and $\tau$-function (\ref{HEtau}) remain invariant while in general case
$\wt\Theta'=\wt\Theta+\frac{1}{2\pi\text{i}}\oint Fdz$.

\section{Conclusion}
The results presented above are generalizable naturally to multicomponent case.
Corresponding systems admit various reductions. In particular, for even
number of components ($2N$), the Hamiltonian reduction,
$$
\mathbf{u}=\sum_{i=1}^N(\Theta_{x_{2i}}\partial_{x_{2i-1}}-\Theta_{x_{2i-1}}\partial_{x_{2i}}),
$$
where $x_i=t^1_i$, $i=1,\dots, 2N$, gives rise to the equations
$$
\Theta_{x_i t_k}-\Theta_{x_k t_i} -\{ \Theta_{x_i},\Theta_{x_k}\}=0,
$$
where $x_i=t^2_i$, $i=1,\dots, 2N$, and
$$
\{F,H\}=\sum_{p=1}^N\left(\frac{\partial F}{\partial x_{2p}}\frac{\partial H}{\partial x_{2p-1}}-
\frac{\partial F}{\partial x_{2p-1}}\frac{\partial H}{\partial x_{2p}}\right).
$$
These equations are associated with linear problems
$$
\mathbf{S}_{t_i}-z\mathbf{S}_{x_i}=\{\mathbf{S},\Theta_{x_i}\}.
$$
The reduction $\mathbf{u}=\sum_{i=1}^{2M}\partial_i +\mathbf{v}$,
where
$$
\mathbf{v}=
\sum_{p=M+1}^N(\Theta_{x_{2p}}\partial_{x_{2p-1}}-\Theta_{x_{2p-1}}\partial_{x_{2p}}),
$$
leads to equations
\be
\Theta_{x_i t_k}-\Theta_{x_k t_i} -\{ \Theta_{x_i},\Theta_{x_k}\}=0,
\label{SDYM}
\ee
where the Poisson bracket is of the form
$$
\{F,H\}=\sum_{p=M+1}^N\left(\frac{\partial F}{\partial x_{2p}}\frac{\partial H}{\partial x_{2p-1}}-
\frac{\partial F}{\partial x_{2p-1}}\frac{\partial H}{\partial x_{2p}}\right).
$$
In the case $N=2$, $M=1$, taking $i=1$, $k=2$ one obtains six-dimensional
generalization of the second heavenly equation \cite{Przanovski}.
The $\dbar$-dressing method for these multidimensional equations will be studied
in elsewhere.
\subsection*{Acknowledgments}
The authors are grateful to S.V. Manakov for fruitful discussions.
LVB was supported in part by RFBR grant 04-01-00508
and President of Russia grant 1716-2003 (scientific schools); BGK was supported in part
by the grant COFIN 2004 `Sintesi'.

\end{document}